\documentclass[aps,twocolumn,prd,showpacs,floatfix]{revtex4-1}

\usepackage{mathrsfs}
\usepackage{dcolumn}
\usepackage{bm}
\usepackage{amsmath}
\usepackage{amsthm}
\usepackage{amssymb}
\usepackage{graphicx}
\usepackage{nicefrac}

\newcommand{\eref}[1]{(\ref{#1})}

\begin{document}

\title{Energy levels of a scalar particle in a static gravitational field
close to the black hole limit}
\author{G. H. Gossel}
\email{g.gossel@student.unsw.edu.au}
\affiliation{School of Physics, University of New South Wales, Sydney 2052, Australia}
\author{J. C. Berengut}
\affiliation{School of Physics, University of New South Wales, Sydney 2052, Australia}
\author{V. V. Flambaum}
\affiliation{School of Physics, University of New South Wales, Sydney 2052, Australia}
\affiliation{European Centre for Theoretical Nuclear Physics (ECT*), Strada della Tabarelle 286, I-38123, Villazzano, Trento, Italy}

\date{17 November 2010}

\begin{abstract}

The bound-state energy levels of a scalar particle in the gravitational field of  finite-sized objects with interiors described by the Florides and Schwarzschild metrics are found. For these metrics, bound states with zero energy (where the binding energy is equal to the mass of the scalar particle) only exist when a singularity occurs in the metric. For the Florides metric this singularity occurs in the black hole limit, while for the constant density (Schwarzschild interior) metric it corresponds to infinite pressure at the center. Moreover, the energy spectrum is shown to become quasi-continuous as the metric becomes singular.

\end{abstract}


\maketitle  

%%%%%%%%%%%%%%%%%%%%%%%%%%%%%%%%%%%%%%%%%%%%%%%%%%%

\section{Introduction}

Many authors have studied quantum phenomena in the gravitational fields of black holes. Quantum scattering events in the black hole background have been considered extensively~\citep{MukhanovWinitzki,Unruh,KerrElectron,Lasenby1,kuchiev}, while explorations of bound states have received somewhat less attention. Quantum bound states in the field of a Schwarzschild black hole have been investigated by Gaina and Ternov using a perturbative approach~\cite{Gaina}. They find quantum analogues of the classical effects of shift of the perihelion of orbit in a central field and precession of the orbital plane in the field of a rotating body. More recent work by Grain and Barrau~\cite{Grain} illustrates that quantum bound states (with finite lifetimes) with no classical analogue exist around black holes. Work on these types of bound states has also been carried out by Lasenby \emph{et~al.}~\citep{Lasenby2}. Perturbative modification of electronic orbitals around charged black holes (black hole ``atoms'') due to the curved manifold has also been considered~\citep{Pindzola}.

Whilst black holes present a physically extreme case, it is not obvious that these interesting quantum effects should be restricted to this type of object; do similar quantum phenomena occur in the fields of finite-sized (non-singular) massive objects?
Two separate but related topics that should be examined in a quantum framework are particle bound states and scattering. The work presented here deals with the former topic; we will consider scattering in a separate paper. 

In this paper we numerically compute ground state energies for a scalar particle in the gravitational field of a non-rotating, spherically symmetric, massive body of finite size described by the Florides interior metric. We examine the behaviour of these states in the limit that the metric becomes singular. This limit is characterized by the approach of the gravitating mass towards the Schwarzschild mass $M_{s}=R/2 G$, where $R$ is the radius of the object and $G$ is the gravitational constant. We also construct an approximate analytic model that verifies the calculated energy eigenvalues and illustrates the behaviour of the bound states. Our results indicate that in the black hole limit the ground state energy tends to zero; that is, the binding energy tends to the rest mass energy of the scalar particle. Additionally, the energy spectrum becomes quasi-continuous in this limit. We repeat our calculations for objects described by the Schwarzschild interior metric and find similar behaviour when this metric becomes singular (in this case this corresponds to a pressure singularity at the origin, rather than the black hole limit).

%%%%%%%%%%%%%%%%%%%%%%%%%%%%%%%%%%%%%%%%%%%%%%%%%%%
\section{Numerical solution of the wave equation}
In order to determine the bound states of a scalar particle in the gravitational field of a finite-sized object we require a wave function that exists for all $r$. This necessitates an interior as well as an exterior metric to substitute into the Klein-Gordon equation (with $\hbar =c=1$)
\begin{equation}
\label{eq:KG}
\partial_\mu (\sqrt{-g} g^{\mu\nu}\partial_\nu \Psi)=\sqrt{-g} m^2\Psi
\end{equation}
where $g^{\mu\nu}$ is the reciprocal of the metric and $m$ is the mass of the scalar particle. The metric on the exterior of a spherically symmetric non-rotating body of mass $M$ is
\begin{equation}
\label{eq:ExteriorMetric}
ds^2 = -\left(1-\frac{r_s}{r}\right)dt^2 +\left(1-\frac{r_s}{r}\right)^{-1}dr^2 + r^2 d\Omega^2
\end{equation}
where $r_s = 2GM$ is the Schwarzschild radius of the gravitating object.
The standard Schwarzschild interior solution for a constant density sphere develops a pressure singularity when $r_s = \frac{8}{9}R$~\citep{Buchdahl}. This forbids the investigation of the limit $r_s \rightarrow R$, thus we cannot use it to examine the black hole case directly. We discuss this solution in Section~\ref{sec:schwarzschild}.

An interior metric that does not contain this singularity is that developed by Florides~\citep{Flor74}:
\begin{equation}
\label{eq:InteriorMetric}
ds^2 = -\frac{\left(1-\frac{r_s}{R}\right)^{\frac{3}{2}}}{\sqrt{1-\frac{r^2 r_s}{R^3}}}dt^2 +\left(1-\frac{r^2 r_s}{R^3}\right)^{-1} dr^2 + r^2 d\Omega^2
\end{equation}
which is valid for $r\leqslant R$. For $r_s=R$
the horizon (coordinate) singularity $r=r_s$~\cite{comment}. Substituting \eref{eq:ExteriorMetric} and \eref{eq:InteriorMetric} into \eref{eq:KG} and assuming a separable solution of the form $\Psi(\vec{r}) = e^{-i\epsilon t}\psi_l(r)Y_{lm}(\theta,\phi)$ we derive the exterior and interior wave equations respectively as
\begin{align}
\label{eq:ExteriorWave}
&\psi''^{E}_{l}(r) + \left(\frac{1}{r-r_s}+\frac{1}{r}\right)\psi'^{E}_{l}(r) +\notag{} \\
& \left(\frac{r^2 \epsilon^2}{(r-r_s)^2}-\frac{m^2 r}{r-r_s}-\frac{l(l+1)}{r(r-r_s)}\right) \psi^{E}_{l}(r) = 0
\end{align}
and
\begin{align}
\label{eq:InteriorWave}
&\psi''^{I}_{l}(r) + \frac{1}{2r}\left(5 -\frac{1}{1 - \frac{r^2 r_s}{R^3}}\right) \psi'^{I}_{l}(r)+\notag{} \\
&\frac{1}{1 - \frac{r^2 r_s}{R^3}}
\left(\frac{(1-\frac{r^2 r_s}{R^3})^{\frac{1}{2}}\epsilon^2}{(1-\frac{r_s}{R})^{\frac{3}{2}}} - m^2 - \frac{l(l+1)}{r^2}\right) \psi^{I}_{l}(r)
=0
\end{align}
with $\epsilon$  the scalar particle energy and $l$ the angular momentum. We see immediately that to have the $\epsilon^2$-dependent term in \eref{eq:InteriorWave} remain finite as $r_s \rightarrow R$ (for $r\neq r_s$), we need $\epsilon/m \sim (1-r_s/R)^{3/4}$. That is, all bound solutions of the Klein-Gordon equation must go towards zero energy in this limit.

The bound states are found by numerically integrating Eqns.~\eref{eq:ExteriorWave} and \eref{eq:InteriorWave} and computing the lowest energy s-wave bound state of the gravitational system for a given $R, m$ and $r_s$. The results are presented as the solid line in Figs.~\ref{fig:smallmu} and \ref{fig:largemu}. Using \eref{eq:ExteriorWave} it is easy to prove that while $r_s < R$ there are no bound state solutions for $m = 0$. However, bound state solutions do exist for any $m>0$.

%%%%%%%%%%%%%%%%%%%%%%%%%%%%%%%%%%%%%%%%%%%%%%%%%%%
\section{Analytic Treatment}
It is instructive to present approximate analytical results which are in good agreement with the numerical solutions. We start with a substitution of the form $\psi(r) = T(r)f(r)$ with
\begin{align}
\label{eq:transformations}
T(r)_I&=\frac{1}{r} \left(1 - \frac{r^2 r_s}{R^3}\right)^{-\nicefrac{1}{8}}\notag{} \\
T(r)_E&=\frac{1}{r} \frac{1}{\sqrt{1-\nicefrac{r_s}{r}}}
\end{align}
for the interior and exterior wave equations, respectively. This transforms equations \eref{eq:ExteriorWave} and \eref{eq:InteriorWave} such that the first derivative of the wavefunction $\psi_l'(r)$ is removed. Additionally, we rescale the energy by the mass such that  $\varepsilon = \epsilon/m$ and all lengths, including the Compton radius of the scalar particle, by the radius of the object $R$: $\rho=r/R$, $s = r_s/R$, $\mu = mR$.
Thus both wave equations are transformed (separately) into the form
\begin{equation}
\label{eq:transformed}
-f''(\rho) + V(\rho) f(\rho) = 0\ .
\end{equation}
%%%%%%%%%%%%%%%%%%%%%%%%%%%%%%%%%%%%%%%%%%%%%%%%%%%
\subsection{Exterior}
Transforming the exterior Klein-Gordon equation \eref{eq:ExteriorWave} yields the following coefficient $V(\rho)$ from \eref{eq:transformed}:
\begin{equation}
\label{eq:transformedexterior}
V_E(\rho) =\frac{l(l+1)}{\rho(\rho-s)}-\frac{s^2}{4\rho^2(\rho-s)^2}-\frac{\varepsilon^2\mu^2\rho^2}{(\rho-s)^2}+\frac{\mu^2\rho}{\rho-s}.
\end{equation}
Taking a series expansion for large $\rho$ to second order yields
\begin{align}
\label{eq:exteriorExpansion}
V_E(\rho)&\approx -\frac{s \mu^2(2\varepsilon^2-1)}{\rho}-\frac{s^2\mu^2(3\varepsilon^2-1)}{\rho^2}\notag{}\\
&+\frac{l(l+1)}{\rho^2}+\mu^2(1-\varepsilon^2).
\end{align}
The weak field limit has been considered previously (see, e.g.~\cite{MukhanovWinitzki} and references within), therefore we will instead concentrate on the strong field limit. We note that free motion corresponds to $V= \mu^2(1-\varepsilon^2)$; that is, $m^2-\epsilon^2$ in the original coordinate system of \eref{eq:ExteriorWave}.

%%%%%%%%%%%%%%%%%%%%%%%%%% 
\subsection{Interior}
Transforming the interior Klein-Gordon equation \eref{eq:InteriorWave} yields the following coefficient $V(\rho)$ from \eref{eq:transformed}:
\begin{align}
V_I(\rho) =
\label{eq:Vi}
& \frac{l(l+1)}{\rho^2} - \frac{7s}{16 (1-\rho^2 s)^2}
  + \frac{\mu^2}{(1-\rho^2 s)} \\
& - \frac{(4l + 5)(4l - 1)s}{16 (1-\rho^2 s)}
  - \frac{\varepsilon^2\mu^2}{(1-s)^{3/2} (1-\rho^2 s)^{1/2}}\notag{}\ .
\end{align}
This corresponds to an effective potential
\[
U(\rho) = \frac{V(\rho)}{2\mu} - \varepsilon^2\mu^2 + \mu^2
\]
where we have subtracted the free motion term $\mu^2(1-\varepsilon^2)$. We can solve Eqs.~(\ref{eq:transformed}), (\ref{eq:Vi}) by expanding around $\rho=0$ to second order in $\rho$ and using the known solutions of the quantum harmonic oscillator. Thus the two equations we wish to compare are
\begin{align*}
V_\textrm{KG}(\rho) =& \frac{l(l+1)}{\rho^2}
  + \left( \mu^2 +l(l+1)s -\frac{3s}{4} 
   - \frac{\varepsilon^2\mu^2}{(1-s)^{3/2}} \right) \\
 &+ \left( \mu^2 s - \frac{19s^2}{16} + l(l+1)s^2
           - \frac{\varepsilon^2 \mu^2 s}{2(1-s)^{3/2}} \right) \rho^2\\
V_\textrm{HO}(\rho) =& \frac{l(l+1)}{\rho^2} + \mu^2\omega^2\rho^2 - 2\mu E
\end{align*}
Equating the coefficients of $r$ and using the solutions of the three dimensional quantum harmonic oscillator, \mbox{$E = (n+\frac{3}{2})\, \omega$} with $n = 2k + l$, $k \geq 0$, gives an equation for the squared scalar particle energy $\varepsilon^2$.
The positive energy solution that corresponds to the energy eigenvalue is
\begin{equation}
\label{eq:epsilon}
\varepsilon = (1-s)^{3/4} (\alpha + \beta)^{1/2}
\end{equation}
where
\begin{align*}
\alpha &= \frac{2n+3}{2\mu^2}
          \sqrt{2\mu^2 s + \left( 2l(l+1) + n(n+3) - 1\right) s^2} \\
\beta &= 1 + \left( l(l+1) - n(n+3) - 3 \right)\frac{s}{\mu^2}\ .
\end{align*}

We test the range of validity of our analytic approximation by considering some physical constraints on our harmonic oscillator model. Firstly, we have assumed that the entire wavefunction is contained within the interior region $\rho \leq 1$ ($r \leq R$). This condition can be satisfied when $\mu$ is large. Formally we require
$\langle \rho^2 \rangle \ll 1$.
$\langle \rho^2 \rangle$ can be obtained using the virial theorem for our harmonic potential
\[
\left< \frac{\mu \omega^2 \rho^2}{2} \right> = \frac{1}{2}(n+3/2)\,\omega\ .
\]
Expanding $\omega$ about $1/\mu$ gives
\begin{gather}
\langle \rho^2 \rangle = \frac{(n+3/2)}{\mu\omega} \ll 1 \nonumber \\
\label{eq:InteriorCondition}
\mu \gg \sqrt{\frac{2}{s}} (n + 3/2)\ .
\end{gather}

Our second test is whether the expansion of $V(\rho)$ as a harmonic potential is valid. The ratio of the coefficient of $\rho^4$ to that of $\rho^2$ is
\[
\frac{5s}{4} + (n+3/2)\sqrt{\frac{s}{2}}\frac{s}{\mu} + O\left(\frac{1}{\mu^2}\right)\ .
\]
If the harmonic oscillator model is appropriate, this should be much less than one for relevant values of $\langle \rho^2 \rangle$. While the coefficient of the $\rho^4$ term clearly begins to dominate at $s\sim 4/5$ (even for large $\mu$), we simply require
\[
\left(\frac{5s}{4} + (n+3/2)\sqrt{\frac{s}{2}}\frac{s}{\mu}\right) \langle \rho^2 \rangle \ll 1
\]
or
\begin{equation}
\mu \gg \frac{5}{4} (n+3/2) \sqrt{2s}\ .
\end{equation}
This is automatically satisfied by the condition \eref{eq:InteriorCondition}.
Coefficients of higher powers of $\rho^2$ in $V(\rho)$ will be suppressed by additional powers of $s$. Furthermore, $\omega > 0$ when \eref{eq:InteriorCondition} is satisfied.
Finally, we note that in the limit $\mu \gg n$, we can expand the energy \eref{eq:epsilon} as
\begin{equation}
\varepsilon = (1-s)^{3/4}\left(1+(n+3/2)\sqrt{\frac{s}{2\mu^2}}+O\left(\frac{s}{\mu^2}\right)\right)\ .
\end{equation}
This is always $\varepsilon < 1$ under condition \eref{eq:InteriorCondition} (the state is bound).
%%%%%%%%%%%%%%%%%%%%%%%%%%%%%%%%%%%%%%%%%%%%%%%%%%%
\section{Summary of Florides Case}
\begin{figure}[t!]
\begin{center}
  \includegraphics[width=0.45\textwidth]{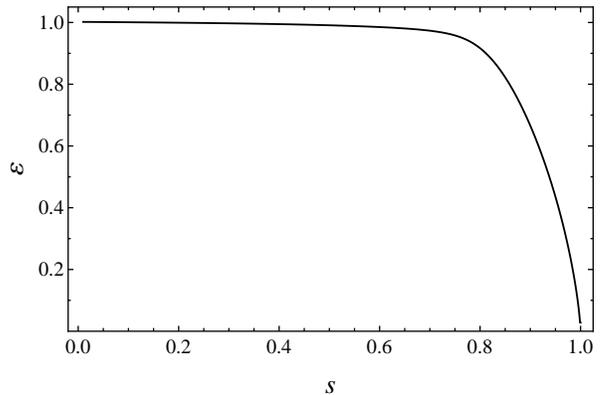}
\caption{\label{fig:smallmu} Numerically computed ground state energy of the Klein-Gordon equation with Florides interior metric for \mbox{$\mu=mR=1/2$}. $s=r_s/R$ is the ratio of the Schwarzschild radius to the radius of the object, and $\varepsilon=\epsilon/m$.}
\end{center}
\end{figure}

\begin{figure}[t!]
\begin{center}
  \includegraphics[width=0.45\textwidth]{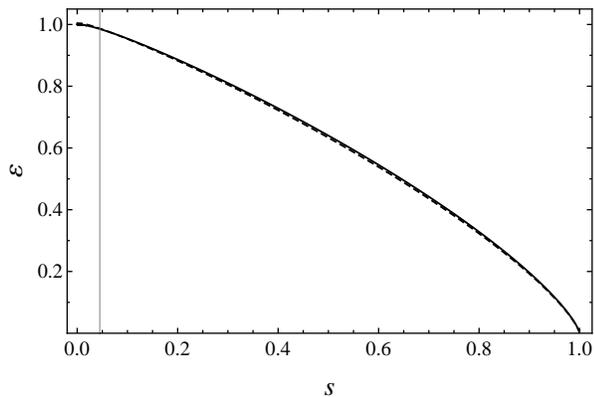}
\caption{\label{fig:largemu} Ground state energy of the Klein-Gordon equation with Florides interior metric in the large $\mu$ regime, \mbox{$\mu=mR=10$}. Solid line: numerically computed solution; dashed line: approximate analytical solution for $\varepsilon=\epsilon/m$ given by Eq.~\eref{eq:epsilon}. The vertical line corresponds to the maximum value of $s=r_s/R$ given by \eref{eq:InteriorCondition} when written as an equality.}
\end{center}
\end{figure}

Figures~\ref{fig:smallmu} and \ref{fig:largemu} depict the ground-state $s$-wave energies obtained by solving the Klein-Gordon equation, \eref{eq:ExteriorWave} and \eref{eq:InteriorWave}, for the small $\mu$ and large $\mu$ regimes, respectively.
In Fig.~\ref{fig:largemu}, one can see that for sufficiently large $\mu$ the analytic and numerically calculated curves are in good agreement. The vertical line in Fig.~\ref{fig:largemu} corresponds to the condition \eref{eq:InteriorCondition} when written as an equality: $s = \sqrt{2}/\mu (3/2) \simeq 0.045$.

From both the numerical and the analytical solutions we find that only in the limit that $r_s \rightarrow R$, is the existence of a bound state with zero energy possible.

Furthermore it can be seen by inspection of \eref{eq:InteriorWave} that in order for the wave equation to be devoid of singular terms as $r_s\rightarrow R$ (for $r \neq r_s$) we require $\epsilon^2\rightarrow 0$, resulting in the entire bound state spectrum becoming quasi-continuous. That is, while all discrete energy levels tend toward zero in this limit, a bound excited state can be found with energy within $d\varepsilon$ of any $\varepsilon>0$ if $d\varepsilon \gtrsim (1-s)^{3/4}$.

%%%%%%%%%%%%%%%%%%%%%%%%%%%%%%%%%%%%%%%%%%%%%%%%%%%
\section{Comment on Schwarzschild Interior}
\label{sec:schwarzschild}
As previously noted, the standard Schwarzschild interior metric cannot be used to calculate $\epsilon$ in the limit $s\rightarrow 1$ as a singularity in the metric occurs at $s=\nicefrac{8}{9}$ where the coefficient of $dt^2$ is zero for $r=0$. Physically this corresponds to the pressure becoming infinite at the center. For completeness we carry out the same numerical analysis of the energy levels and compare the result to that obtained by considering the particle in Florides metric. The interior solution for a constant density fluid sphere developed by Schwarzschild~\cite{Schwarzschild} is given by
\begin{align}
\label{eq:SchwarzschildInterior}
ds^2 =&-\left(\frac{3}{2}\sqrt{1-\frac{r_s}{R}}-\frac{1}{2}\sqrt{1-\frac{r_s r^2}{R^3}}\right)^2 dt^2\notag \\
&+\left(1-\frac{r_s r^2}{R^3}\right)^{-1}dr^2 +r^2 d\Omega^2.
\end{align}
\begin{figure}[b!]
\begin{center}
  \includegraphics[width=0.45\textwidth]{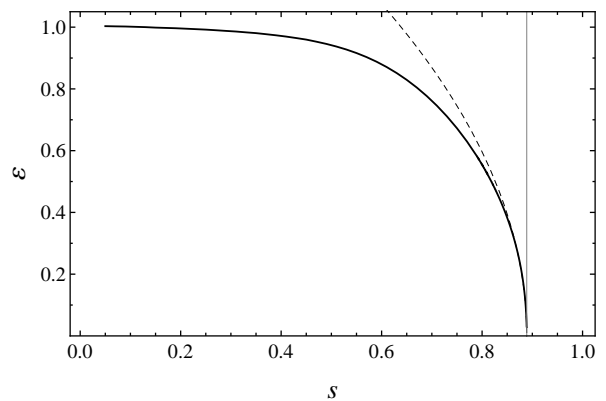}
\caption{\label{fig:SMetricFig} Ground state energy of the Klein-Gordon equation for the Schwarzschild interior metric with $\mu=1$. Solid line: numerically computed solution; dashed line: analytical solution given by \eref{eq:Delta1} with $C=2$. The vertical line corresponds to $s=8/9$.}
\end{center}
\end{figure}
Following the same procedure used to generate \eref{eq:ExteriorWave} and \eref{eq:InteriorWave} we construct an interior wave equation for $\psi(r)$.  After rescaling the system as we did in the Florides case this wave equation is solved numerically, the results of which are depicted in Fig.~\ref{fig:SMetricFig} as the solid line.

In order to analytically approximate $\varepsilon$ in this system we transform the wave equation
\begin{equation}
\label{eq:SInteriorWave}
-\psi''(\rho)+A\psi'(\rho)+B\psi(\rho)=0
\end{equation}
\begin{align*}
A=&\frac{s\rho(s(9-\rho^2)-8)}{2(1-s\rho^2)\left(\sqrt{1-s\rho^2}-3\sqrt{1-s}\right)^2}-\frac{2}{\rho(1-s\rho^2)}\\
&+\frac{21s\rho}{6(1-s\rho^2)},
\\
B=&\frac{\mu^2}{(1-s\rho^2)}\left(1+\frac{4\varepsilon^2}{\left(\sqrt{1-s\rho^2}-3\sqrt{1-s}\right)^2}\right)
\end{align*}
into an effective potential form. Firstly we define new the parameter
 $~\delta=\frac{8}{9}-s$ and new coordinate $~ x=\frac{r}{9R}\sqrt{\frac{8}{\delta}}$. After taking series expansion to order zero in $\delta$, and performing the transformation
\begin{equation*}
\psi_I(x)=T(x)f(x)=\frac{1}{x}\frac{1}{\sqrt{x^2+1}}f(x)
\end{equation*}
we generate a wave equation for $f(x)$ of the form
\begin{equation}
-f''(x)+V(x)f(x)=0
\end{equation}
where
\begin{align}
\label{eq:SchwarzschildVEff}
V(x)&=\frac{2x^2+3}{(x^2+1)^2}-\frac{2\varepsilon^2\mu^2}{\delta(x^2+1)^2}-\frac{9\varepsilon^2\mu^2(x^2+2)(2x^2+1)}{2(x^2+1)^3}.
\end{align}
In order for \eref{eq:SchwarzschildVEff} to remain finite for a given $\mu$ in the limit that $\delta\rightarrow0$ we must have 
\begin{equation}
\label{eq:Delta1}
\varepsilon= C\frac{\sqrt{\delta}}{\mu}= C\frac{\sqrt{8/9-s}}{\mu}
\end{equation}
where $C$ is a proportionality constant of order $1$. To clarify, in our rescaled coordinates $\langle x^2\rangle$ is independent of delta. Thus the condition on $\epsilon$ as $\delta$ becomes small is valid. Eqn.~\eref{eq:Delta1} is plotted for $C=2$ and $\mu=1$  as the dashed line in Fig.~\ref{fig:SMetricFig}.

The range of $\delta$ where this approximation is valid is found by considering the relative size of higher order $\delta$ terms in the series expansion used to generate \eref{eq:SchwarzschildVEff}. By requiring subsequent higher order $\delta$ terms to be small compared to \eref{eq:SchwarzschildVEff}, we find the range of validity of \eref{eq:Delta1} for $\mu\gg1$ is
\begin{equation}
\label{eq:SchwarzschildCon1}
\delta\ll\frac{1}{10\mu^2}.
\end{equation}
The energy is small ($\varepsilon\ll1$) if
\begin{equation}
\label{eq:SchwarzschildCon2}
\delta\ll\mu^2.
\end{equation}
 Together these constraints provide an adequate estimate of the window of $\delta$ for which \eref{eq:Delta1} is valid.

From this analysis we find that the energy levels of scalar particles bound in the field of the Schwarzschild interior metric approach zero as the metric becomes singular at $s=\nicefrac{8}{9}$.

%%%%%%%%%%%%%%%%%%%%%%%%%%%%%%%%%%%%%%%%%%%%%%%%%%%
\section{Discussion}
By matching the Florides interior metric to the standard Schwarzschild exterior metric we are able to use the Klein-Gordon equation to construct the wave function over all space for a scalar particle in the gravitational field of a finite-sized spherically-symmetric static mass. Numerical computation of the ground state energies of the scalar particle shows that the existence of a zero energy bound state may only be possible in the limit that the gravitating object described by the Florides metric becomes a black hole.

This result is verified using an approximate analytical calculation.
 It is seen that the entire bound state spectrum collapses to $\epsilon = 0$ as $r_s\rightarrow R$. The spectrum becomes quasi-continuous in this limit: the 
energies and the intervals between the energy levels are proportional to
$(1-r_s/R)^{3/4}$. If we keep energy fixed the principal quantum number of the level $n$ tends to infinity. Similar results are obtained by considering the behaviour of a scalar particle in the background of a fluid sphere of constant density as modeled by the Schwarzschild interior metric. In this case we find that as the metric becomes singular at $s=r_s/R=8/9$ the energy of the particle tends to zero as
 $\varepsilon \propto \frac{1}{\mu}\sqrt{8/9-s}$.

The existence of a bound state with zero energy is relevant to the phenomenon of particle pair production. Due to quantum fluctuations, particle anti-particle pairs are produced around a black hole and it is possible that one of the pair escapes the gravitational field: this is known as the Hawking radiation~\cite{Hawking74, Hawking75}. This occurs when the pair is created with each member of the pair on opposite sides of the event horizon: one particle falls in while the other escapes to infinity. 

Other systems that give rise to pair production, also known as vacuum breakdown, include static and dynamic electromagnetic fields (see, e.g.~\cite{Ruffini08}) and time-varying gravitational fields (see, e.g.~\cite{Hossenfelder}). In the Coulomb case it is clear that one of the pair must be repelled, since the pair have opposite charge. In the gravitational case, however, both particles will possess the same `gravitational charge' and thus will be attracted by the potential. In the black hole case the existence of the event horizon as a barrier facilitates one particle escaping to infinity as the other falls to the singularity.

For zero energy states to exist (a necessary but not sufficient condition for particle production and Hawking radiation) in the metrics we have considered, the metric must become singular. For pair production with the ejection of one particle to infinity it is actually necessary to have a negative energy level $\epsilon < -mc^2$. It is instructive to compare the gravitational field case with the electrostatic potential case where pair creation is possible in a strong Coulomb field $U(r)$. (For example, the ground state of an electron orbiting a finite-size nucleus reaches the lower continuum when $Z \gtrsim 170$. For a scalar particle of the same mass, $Z$ should be greater than $\sim 90$~\cite{popov71sjnp}.) In the Coulomb case we have  $(\epsilon - U(r))^2$ in the wave equation. Therefore, increasing the absolute value $|U|$ of the  negative potential $U(r)$ leads to the negative energy $\epsilon$. In the gravitational field we have  $\epsilon^2$ in the wave equation and such a procedure cannot introduce negative energy bound states.\\

\acknowledgments

We thank M. Yu. Kuchiev and G. F. Gribakin for useful discussions. This work is supported by the Australian Research Council and ECT*.


\begin{thebibliography}{99}
\bibitem{comment} To avoid misunderstanding, we should note that this metric
does not correspond to any macroscopic object for $R < 3r_s/2$. This is further
discussed by N. K. Kofinti, Gen. Rel. Grav. {\bf 17}, 245 (1985).
\end{thebibliography}
\end{document}